\documentclass[english]{article}
\usepackage[T1]{fontenc}
\usepackage[latin9]{inputenc}
\usepackage{geometry}
\usepackage{mathrsfs}
\usepackage{amsmath}
\usepackage{amssymb}
\usepackage{graphicx}
\usepackage{setspace}
\usepackage{esint}
\makeatletter

\providecommand{\tabularnewline}{\\}

\date{}

\makeatother

\usepackage{babel}
\begin{document}

\title{Bayesian analysis of multivariate stable distributions using one-dimensional
projections}

\author{Mike G. Tsionas\thanks{Economics Department, Lancaster University Management School, LA1
4YX UK email: $\mathsf{m.tsionas@lancaster.ac.uk}$}}
\maketitle
\begin{abstract}
In this paper we take up Bayesian inference in general multivariate
stable distributions. We exploit the representation of Matsui and
Takemura (2009) for univariate projections, and the representation
of the distributions in terms of their spectral measure. We present
efficient MCMC schemes to perform the computations when the spectral
measure is approximated discretely or, as we propose, by a normal
distribution. Appropriate latent variables are introduced to implement
MCMC. In relation to the discrete approximation, we propose efficient
computational schemes based on the characteristic function.\end{abstract}
\begin{quote}
\end{quote}
\begin{quotation}
Acknowledgments: The paper benefited from the comments of two anonymous
referees.

\pagebreak{}
\end{quotation}

\section{Introduction}

Univariate stable distributions have been thoroughly studied in econometrics,
statistics and finance over the past few decades (Samorodnitsky and
Taqqu, 1994). Their empirical application is still hampered by the
fact that their density is not available in analytic form, despite
advances in Bayesian computation using MCMC. Buckle (1995) and Tsionas
(1999) provided Gibbs sampling schemes for general and symmetric stable
distributions, respectively. The problem is that the conditional posterior
distributions of certain latent variables are cumbersome to work with
and require careful tuning. The analogous problem in the multivariate
case is exceedingly difficult although a few attempts have been made
to solve it. The impediment is that multivariate stable distributions,
unlike the univariate case, are defined through their spectral measures
which, in practice, are unknown. Ravishanker and Qiou (1999) for example,
proposed an EM algorithm based on Buckle (1995) in the case of symmetric
isotropic stable distributions but this class is too narrow to be
of empirical importance. It is defined by the transformation $X=\mu+C\xi$,
where $\xi$ is a vector of independent random variables each one
distributed as standard symmetric stable, $\mu$ is a vector of location
parameters, $\Sigma$ is a scale matrix, and $C^{\top}C=\Sigma$.
It is known that the class of elliptical stable distributions can
be defined through the transformation $X=\mu+RCu$ where $u$ is uniformly
distributed on the unit sphere $\mathbb{S}^{d-1}=\{\boldsymbol{x}\in\mathbb{R}^{d}|\:||\boldsymbol{x}||\boldsymbol{}=1\}$,
$C$ is a $d\times d$ scale matrix of full rank, and $R=\sqrt{VS_{\alpha/2}}$
where, independently, $V\sim\chi_{d}^{2}$ and $S_{\alpha/2}$ follows
a stable distribution with parameter $\alpha/2$ and maximal skewness
$\beta=1$. Of course not all multivariate stable distributions are
elliptical. See Lombardi and Veredas (2009). When $V\sim\chi_{1}^{2}$
the distribution of $X$ is in the class of elliptically contoured
stable distributions (Nolan, 2006, p.2). 

In connection with multivariate stable Paretian distributions, even
the computation of the characteristic functions becomes complicated
because they are only defined through their spectral measure, an object
that is needed in order to retain the equivalence between the density
and the characteristic function. The estimation of the spectral measure
itself has proved itself to be quite cumbersome even for bivariate
distributions (see the seminal works of McCulloch, 1994, 2000, Nolan
et al., 2001, and Nolan and Rajput, 1995). 

The present paper is related to recent advances in the econometrics
of stable distributions. Dominicy and Veredas (2012) propose a method
of quantiles to fit symmetric stable distributions. Since the quantiles
are not available in closed form they are obtained using simulation
resulting in the method of simulated quantiles or MSQ. Hallin, Swan,
Verdebout and Veredas (2012) propose an easy-to-implement R-estimation
procedure which remains -consistent contrary to least squares with
stable disturbances. Broda, Haas, Krause, Paolella and Steude (2012)
propose a new stable mixture GARCH model that encompasses several
alternatives and can be extended easily to the multivariate asset
returns case using independent components analysis. Ogata (2012) uses
a discrete approximation to the spectral measure of multivariate stable
distributions and proposes estimating the parameters by equating the
theoretical and empirical characteristic function in a generalized
empirical likelihood / GMM framework. 

Relative to this work, we show how to implement Bayesian inference
for multivariate stable distributions by providing statistical inferences
about the spectral measure jointly with the other parameters of the
model. For numerical analysis via MCMC we employ a novel data augmentation
technique for stable distributions. We use a discrete approximation
of the measure where the configuration and the number of points are
unknown. We also propose a novel approximation to the spectral measure
based on a multivariate normal distribution.

\section{Stable distributions}

A random variable $X$ is called strictly (univariate) stable if for
all $n$, $\sum_{i=1}^{n}X_{i}\sim c_{n}X$ for some constant $c_{n},$
where $X_{1},...,X_{n}$ are independently distributed with the same
distribution as $X$. It is known that the only possible choice is
to have $c_{n}=n^{1/\alpha}$ for some $\alpha\in(0,2]$. General
non-symmetric stable distributions are defined via the log characteristic
function which is given by the following expression (Samorodnitsky
and Taqqu, 1994, and Zolotarev, 1986):

\begin{equation}
\begin{array}{c}
\log\varphi(\tau)=\log E\exp(\iota\tau X)=\\
\begin{cases}
\begin{array}{c}
\iota\mu\tau-|\sigma\tau|^{\alpha}\{1-\iota\beta\mathrm{sgn}(\tau)\tan\frac{\pi\alpha}{2}\},\;\alpha\neq1\\
\iota\mu\tau-\sigma|\tau|\{1+\iota\beta\mathrm{sgn}(\tau)\frac{2}{\pi}\log|\tau|\},\;\alpha=1,
\end{array}\end{cases}
\end{array}
\end{equation}

where $\tau\in\mathbb{R}$, $\mu$ and $\sigma$ are location and
scale parameters, $\alpha$ is the characteristic exponent, $\beta\in[-1,1]$
is the skewness parameter, and $\iota=\sqrt{-1}.$ In this paper we
are interested in multivariate stable distributions, that is distributions
of a random variable in $\mathbb{R}^{d}$. Suppose $X$ is a vector
of random variables with characteristic exponent $\alpha\in(0,2]$.
Its characteristic function is $\varphi_{X}(\boldsymbol{\tau})=E\exp\{\iota<\boldsymbol{\tau},X>)=\exp\left(-I_{X}(\boldsymbol{\tau})+\iota<\tau,\mu>\right)$
where $<\boldsymbol{\tau},X>=\boldsymbol{\tau}^{\top}X$ denotes inner
product, and

\begin{equation}
I_{X}(\boldsymbol{\tau})=\int_{\mathbb{S}^{d-1}}\psi_{\alpha}\left(<\boldsymbol{\tau},\boldsymbol{s}>\right)\Gamma(d\boldsymbol{s}),
\end{equation}

where $\mathbb{S}^{d-1}=\left\{ \boldsymbol{u}\in\mathbb{R}^{d}|<\boldsymbol{u},\boldsymbol{u}>=1\right\} $
is the boundary of the unit ball in $\mathbb{R}^{d}$, $\Gamma$ is
a finite Borel measure of the vector $X$, called the spectral measure,
$\mu\in\mathbb{R}^{d}$ is a vector of location parameters, and the
complex function $\psi$ is defined as follows:

\begin{equation}
\psi_{\alpha}(u)=\begin{array}{c}
\begin{cases}
|u|^{\alpha}\{1-\iota\mathrm{sgn}(u)\tan\frac{\pi\alpha}{2}\}, & \;\alpha\neq1,\end{cases}\\
\begin{cases}
|u|\{1+\iota\frac{2}{\pi}\mathrm{sgn}(u)\log|u|\}, & \;\alpha=1\end{cases}.
\end{array}
\end{equation}

See seminal work by Nolan (1998), Nolan and Rajput (1995), Abdul-Hamid
and Nolan (1998), and also Cambanis and Miller (1981), and Nagaev
(2000). Notably the parameters $(\alpha,\Gamma)$ fully define all
centered multivariate stable distributions, and a skewness parameter
$\beta$ is not needed\footnote{Actually, there are skewness parameters $\beta(\boldsymbol{\tau})$
that depend on the particular projection $\boldsymbol{\tau}$.} in this case, since we have the full measure, $\Gamma$. We denote
the class by $X\sim\mathcal{\mathscr{S}}_{\alpha,d}(\mu,\Gamma)$.
Press (1972) attempted to define a multivariate $\alpha$-stable distribution
without using the spectral measure $\Gamma$. Later on Paulauskas
(1976) provided some corrections as not all $\alpha$-stable distributions
can be represented using Press\textquoteright{} (1972) characteristic
function. Chen and Rachev (1995) is an interesting paper where the
authors provided estimates of the spectral measure as well as applications
to a stable portfolio. It is notable that the projection of $X$ on
$\tau,$ viz. $<\boldsymbol{\tau},X>$ has a univariate stable distribution.
The characteristic exponent $\alpha$ remains the same but scale,
location and skewness depend on $\boldsymbol{\tau}$.The multivariate
characteristic function is not easy to work with as in the univariate
case because of the dependence on the spectral measure. As this can
rarely be specified in advance, it is necessary to provide posterior
inferences about it, in the context of Bayesian analysis. 

One approach (Byczkowski et al., 1993) is to assume that $\Gamma$
can be approximated by a discrete measure, in which case we have:

\begin{equation}
\Gamma(d\boldsymbol{s})=\sum_{j=1}^{J}\gamma_{j}\delta_{\{\boldsymbol{s}^{(j)}\}}(d\boldsymbol{s}),
\end{equation}

where $\gamma_{j}>0$, $\boldsymbol{s}^{(j)}\in\mathbb{S}^{d},j=1,...,J$,
$\delta$ denotes Dirac's function and the approximation is made at
$J$ points of the unit sphere in $\mathbb{R}^{d}$. The meaning of
(4) is that we define a finite partition $A_{1},...,A_{J}$ of $\mathbb{S}^{d-1}$,
points $\boldsymbol{s}^{(1)},...,\boldsymbol{s}^{(J)}\in\mathbb{S}^{d-1}$
and construct $\Gamma$ by placing mass $\Gamma(A_{j})$ at $\boldsymbol{s}^{(j)}$
so that:

\[
\Gamma(d\boldsymbol{s})=\sum_{j=1}^{J}\Gamma(A_{j})\delta_{\{\boldsymbol{s}^{(j)}\}}(d\boldsymbol{s}).
\]

Suppose $\theta$ denotes the parameters. Since $\varphi_{X}(\boldsymbol{\tau};\theta)=\exp\left\{ -\intop_{\mathbb{S}^{d}}\psi_{\alpha}(<\boldsymbol{\tau},\boldsymbol{s}>)\Gamma(d\boldsymbol{s})\right\} $
we obtain:

\begin{equation}
\varphi_{X}(\boldsymbol{\tau};\theta)=\exp\left(-\sum_{j=1}^{J}\gamma_{j}\psi_{\alpha}(<\boldsymbol{\tau},\boldsymbol{s}^{(j)}>)\right),
\end{equation}

or alternatively:

\begin{equation}
-\log\varphi_{X}(\boldsymbol{\tau}^{(i)},\boldsymbol{s}^{J};\theta)=\sum_{j=1}^{J}\gamma_{j}\psi_{\alpha}(<\boldsymbol{\tau},\boldsymbol{s}^{(j)}>),i=1,...,J,
\end{equation}

where $\boldsymbol{\tau}^{K}=(\boldsymbol{\tau}^{(1)},...,\boldsymbol{\tau}^{(J)})$
denotes a set of points where the log-characteristic function is evaluated
and $\boldsymbol{s}^{J}=(\boldsymbol{s}^{(1)},...,\boldsymbol{s}^{(J)})$.
If we define:

\[
\begin{array}{c}
\mathbb{Y}(\boldsymbol{\tau}^{K},\boldsymbol{s}^{J})=[-\log\varphi_{X}(\boldsymbol{\tau}^{(1)},\boldsymbol{s}^{J};\theta),...,-\log\varphi_{X}(\boldsymbol{\tau}^{(K)},\boldsymbol{s}^{J};\theta)]'=\\
{}[I_{X}(\boldsymbol{\tau}^{(1)},\boldsymbol{s}^{J};\theta),...,I_{X}(\boldsymbol{\tau}^{(K)},\boldsymbol{s}^{J};\theta)]'\\
\mathbb{X}(\boldsymbol{\tau}^{K},\boldsymbol{s}^{J})=[\mathscr{X}_{ij}(\boldsymbol{\tau}^{(i)},\boldsymbol{s}^{(j)})]\\
\mathscr{X}_{ij}(\boldsymbol{\tau}^{(i)},\boldsymbol{s}^{(j)})=\psi_{\alpha}(<\boldsymbol{\tau}^{(i)},\boldsymbol{s}^{(j)}>),i=1,...,K,j=1,...,J,
\end{array}
\]

$ $

$ $we can write (6) as a system of linear equations\footnote{The system is in general complex-valued so we need to take the real
and imaginary parts of $\mathbb{Y}$ and $\mathbb{X}$. }:

\label{CF system no error}
\begin{equation}
\mathbb{Y}(\boldsymbol{\tau}^{K},\boldsymbol{s}^{J})=\mathbb{X}(\boldsymbol{\tau}^{K},\boldsymbol{s}^{J})\gamma,
\end{equation}

from which, in principle, we can obtain approximations to the spectral
weights, $\gamma_{j}$ which, in (7), we collect in vector $\gamma$.
In practice, the system of equations suffers from singularities and
the estimates of $\gamma$ are not always non-negative. The reason
that $\mathbb{X}^{\top}\mathbb{X}$ is often singular is that when
a full grid is used, we encounter points where $I_{X}(-\boldsymbol{\tau})=\overline{I_{X}(\boldsymbol{\tau})}$.
Nolan, Panorska and McCulloch (2001) propose such symmetric grids
around the basic directions (1,0), (0,1), (-1,0), and (0,-1) corresponding
to independent components in the bivariate case. Generally we would
need $J\propto2^{d}$ in the $d-$dimensional case \emph{if we need}
to explore the measure around coordinates corresponding to independent
components. This is manageable in dimensions up to 10. Of course the
possibility arises that we \emph{may} actually need a value of $J$
that is \emph{significantly lower }since the spectral measure can
be embedded in a much smaller subspace. McCulloch (1994, 2000) has
proposed the use of quadratic programming imposing the non-negativity
and Nolan, Panorska and McCulloch (1997) report that, at least in
small dimensions, the procedure works well. The problem is challenging
in that a double grid has to be specified, $\boldsymbol{\tau}^{K}$
for the set of points to evaluate the log-characteristic function
and $\boldsymbol{s}^{J}$ for the support of the discrete measure.
Apparently in all but very low dimensions ($d=2$ specifically) if
we were to use a full grid in (7) we would face the curse of dimensionality
as matrix $\mathbb{X}$ would be huge. Therefore, the only choice
is to place the points $\boldsymbol{\tau}^{K},\boldsymbol{s}^{J}$
in a ``wise'' manner without sacrificing computational ease.

\section{A hierarchical model for multivariate stable distributions}

Matsui and Takemura (2009) extended the work of Abdul-Hamid and Nolan
(1998) and provided semi-closed expressions for the class of multivariate
stable distributions. Abdul-Hamid and Nolan (1998) used higher order
derivatives of one-dimensional densities. The role of the following
function is important:
\begin{quote}
\begin{alignat}{1}
g_{\alpha,d}(v,\beta) & =\begin{cases}
\begin{array}{c}
(2\pi)^{-d}\intop_{0}^{\infty}\cos\left(vu-\left\{ \beta\tan\frac{\pi\alpha}{2}\right\} u^{\alpha}\right)u^{d-1}\mathrm{exp}(-u^{\alpha})du,\;\alpha\neq1,\\
(2\pi)^{-d}\intop_{0}^{\infty}\cos\left(vu+\tfrac{2}{\pi}\beta u\log u\right)u^{d-1}\mathrm{exp}(-u)du,\;\alpha=1.
\end{array}\end{cases}
\end{alignat}

\end{quote}
By theorem 1.1 in Matsui and Takemura (2009) due to Theorem 1 in Nolan
(1998), the density of $X\sim\mathscr{S}_{\alpha,d}(\zeta,\Gamma)$,
where $\zeta\in\mathbb{R}^{d}$ is a shift parameter, can be expressed
as:
\begin{quote}
\begin{alignat}{1}
f_{\alpha,d}(x) & =\begin{cases}
\begin{array}{c}
\int_{\mathbb{S}^{d-1}}g_{\alpha,d}\left(\frac{<x-\zeta,\boldsymbol{S}>}{\sigma(\boldsymbol{S})},\beta(\boldsymbol{S})\right)\sigma(\boldsymbol{S})^{-d}d\boldsymbol{S},\;\alpha\neq1,\\
\int_{\mathbb{S}^{d-1}}g_{1,d}\left(\frac{<x-\zeta,\boldsymbol{S}>-\mu(\boldsymbol{S})-\tfrac{2}{\pi}\beta(\boldsymbol{S})\sigma(\boldsymbol{S})\log\sigma(\boldsymbol{S})}{\sigma(\boldsymbol{S})},\beta(\boldsymbol{S})\right)\sigma(\boldsymbol{S})^{-d}d\boldsymbol{S},\;\alpha=1.
\end{array}\end{cases}
\end{alignat}

\end{quote}
Here, the parameters 

\label{sigma(S)}
\begin{equation}
\sigma(\boldsymbol{S})=\left(\int_{\mathbb{S}^{d-1}}|<\boldsymbol{t},\boldsymbol{S}>|^{\alpha}\Gamma(d\boldsymbol{t)}\right)^{1/\alpha},
\end{equation}

\label{beta(S)}
\begin{equation}
\beta(\boldsymbol{S})=\sigma(\boldsymbol{S})^{-\alpha}\int_{\mathbb{S}^{d-1}}\mathrm{sgn}(<\boldsymbol{t},\boldsymbol{S}>)\cdot|<\boldsymbol{t},\boldsymbol{S}>|^{\alpha}\Gamma(d\boldsymbol{t}),
\end{equation}

$ $ and in these expressions $\boldsymbol{t}\in\mathbb{R}^{d}$ is
the angle, see Samorodnitsky Taqqu (1994, example 2.3.4) and Matsui
and Takemura (2009). Notably, the location or shift parameter is zero
when $\alpha\neq1$ and a certain constant independent of $\zeta$
when $\alpha=1$. We can certainly assume $\zeta=0$. The role of
$\zeta$ arises mainly when the distribution is symmetric around $\zeta$,
see Corollary 4 in Abdul-Hamid and Nolan (1998). For the definition
of location when $\alpha=1$ see equation (6) in Matsui and Takemura
(2009) or equation (9) in Abdul-Hamid and Nolan (1998). The advantage
of the expressions is that function $g_{\alpha,d}$ ``is a function
of two real variables no matter what the dimension $d$ is, and that
this function is the same for every $\alpha$-stable r.v. $X$; i.e.,
it is independent of the spectral measure'' (Abdul-Hamid and Nolan,
1998).

It is possible to represent the density in terms of functions similar
to those used by Zolotarev (1986, equation 2.2.18, p.74 ) to convert
the range of integration to a finite interval (whose upper bound is
one) and avoid the infinite oscillations caused by the trigonometric
terms (Buckle, 1995). It is clear, however, that latent variables
$\boldsymbol{s}$ can be defined so that when $X_{t}\sim\mathrm{iid}\mathscr{S}_{\alpha,d}(\zeta,\Gamma)$,
$t=1,...,n$, then from (8) and (9) we have:

\begin{equation}
p\left(X_{t}|\boldsymbol{S}_{t},\theta\right)\propto g_{\alpha,d}\left(v_{t};\beta(\boldsymbol{S}_{t})\right),t=1,...,n,
\end{equation}

\begin{equation}
v_{t}\equiv\frac{<X_{t}-\zeta,\boldsymbol{S}_{t}>}{\sigma(\boldsymbol{S}_{t})},
\end{equation}

\begin{equation}
\boldsymbol{S}_{t}\sim\mathscr{U}(\mathbb{S}^{d-1}),t=1,...,n,
\end{equation}

and $\mathscr{U}(\mathbb{S}^{d-1})$ denotes the uniform distribution
over the unit sphere. We remark that:

\begin{equation}
\sigma(\boldsymbol{S}_{t})=\left(\intop_{\mathbb{S}^{d-1}}|<\boldsymbol{S}_{t},\boldsymbol{c}>|^{\alpha}\Gamma(d\boldsymbol{c})\right)^{1/\alpha},
\end{equation}

\begin{equation}
\beta(\boldsymbol{S}_{t})=\sigma(\boldsymbol{S}_{t})^{-\alpha}\intop_{\mathbb{S}^{d-1}}\mathrm{sgn}(<\boldsymbol{S}_{t},\boldsymbol{c}>)|<\boldsymbol{S}_{t},\boldsymbol{c}>|^{\alpha}\Gamma(d\boldsymbol{c})
\end{equation}

The representation in (10)-(12) is a hierarchical model for a general
multivariate stable distribution. Of course, the hierarchical model
involves the spectral measure, $\Gamma(d\mathbf{\boldsymbol{S}})$,
through the functions $\beta(\boldsymbol{S})$ and $\sigma(\boldsymbol{S})$.
These functions can be computed either via direct integration over
the unit sphere in $\mathbb{R}^{d}$ or by simulation (when $\boldsymbol{S}\sim\Gamma$
in their computation). It is possible to use further augmentation
of the model by latent variables which is, however, not recommended
since it will affect seriously the mixing properties of MCMC. See
for example the integrals over a bounded interval used by Matsui and
Takemura (2009) to obtain the $g_{\alpha,d}$ function. These correspond
to similar expressions in Zolotarev (1986, pp. 76-77) and Buckle (1995).
Although numerical integration is facilitated it does not avoid the
problem of the presence of the spectral measure, $\Gamma$. Eventually,
the density (9) has to be evaluated using $v_{t}$ in (14) and as
Matsui and Takemura (2009) note the singularities are not completely
avoided and further work is needed to deal with them as in section
3 of Nolan (1997) or the work reported here.

Computation of (8) is a subtle matter. We proceed as follows. Since
the cosine function $\cos(x)=0$ at $x=\frac{k\pi}{2}$, for $k=0,\frac{1}{2},1,\frac{3}{2},2,...$
we locate the roots of the equation $vu-\left\{ \beta(\boldsymbol{s}_{t})\tan\frac{\pi\alpha}{2}\right\} u^{\alpha}=\frac{k\pi}{2}$,
for $k=1,\frac{3}{2},2,...$ with respect to $u$. Denote these roots
by $r_{1},r_{2},...$. The largest value of $k$ is determined by
the weight factor in (8), viz. $f(u)=u^{d-1}\mathrm{exp}(-u^{\alpha})du$
so that the value of the weight factor at a certain $\bar{u}$ is
less than $\varepsilon=10^{-7}$ relative to its mode. The weight
factor can be modified using the transformation $w=u^{\alpha}$ in
which case $w$ follows a standard \emph{gamma} distribution with
shape parameter $\frac{d}{\alpha}$. The mode is $m=\left(\frac{d}{\alpha}-1\right)^{1/\alpha}$
so in practice we can set $\bar{u}=Cm$ and determine the constant
$C$ so that the weight factor is sufficiently small. Therefore the
integral in (15) is computed in the intervals $[0,r_{1}]$, $[r_{1},r_{2}]$,
etc as follows:

\begin{equation}
\begin{array}{c}
g_{\alpha,d}(v,\beta(\boldsymbol{S}_{t}))=\\
(2\pi)^{-d}\sum_{k=1}^{I}\intop_{r_{k-1}}^{r_{k}}\cos\left(vw^{1/\alpha}-\left\{ \beta(\boldsymbol{S}_{t})\tan\frac{\pi\alpha}{2}\right\} w\right)w^{\frac{d}{\alpha}-1}\mathrm{exp}(-w)dw
\end{array}
\end{equation}

after a change of variables to $w=u^{\alpha},$ where $r_{0}=0$ and
$r_{I}>\bar{w}=10\left(\frac{d}{\alpha}-1\right)$. This is always
possible since the cosine function has an infinite number of roots.
We use 20-point Gaussian quadrature to compute the integrals in the
intervals $[r_{k-1},r_{k}]$ determined by the roots of $\chi_{k}(u)=vu-\left\{ \beta(\boldsymbol{s}_{t})\tan\frac{\pi\alpha}{2}\right\} u^{\alpha}-\frac{k\pi}{2}=0$.
An alternative stopping criterion we used is when the roots $r_{I}-r_{I-1}<\epsilon=10^{-4}$
so that the contribution to the integral in (8) is trivial. We have
found that locating the roots is extremely easy provided we can locate
the root of $\chi_{1}(u)=0$ which requires a good starting value.
Then the root $r_{k}$ is an excellent starting value to locate $r_{k+1}$
using a standard Newton algorithm with analytical derivatives. The
root $r_{1}$, viz. $\chi_{1}(r_{1})=0$ can be located using bisection.
Moreover, Gaussian quadrature was found to work well. This procedure
takes account of the oscillations of $g_{\alpha,d}(v,\beta(\boldsymbol{s}_{t}))$
and is quite efficient in computing its values. In practice, too many
zero points are needed and, therefore, it is better to utilize the
finite integral representation in Matsui and Takemura (2009) and Abdul-Hamid
and Nolan (1986). Our results in artificial and actual data were compared
with 30- and 40-point quadrature and we found no essential differences.
We were not able to find different results when the finite integral
representation mentioned above was used. However, the form of the
integrand suggests an adaptive quadrature scheme in a finite interval,
until specified tolerance is achieved. This method is comparable in
terms of accuracy, but less efficient compared to Matsui and Takemura
(2009) and Abdul-Hamid and Nolan (1986).

Here, \emph{we propose to approximate the spectral measure} \emph{by
\[
\boldsymbol{c}_{t}^{*}\sim\mathscr{N}_{d}(\mu\cdot1_{d},\omega^{2}I),\:\boldsymbol{c}_{t}=\frac{\boldsymbol{c}_{t}^{*}}{|\boldsymbol{c}_{t}^{*}|},
\]
}

viz. \emph{a multivariate normal over the unit sphere}, where $\mu$
and $\omega$ are unknown parameters and $1_{d}$ is the unit vector
in $\mathbb{R}^{d}$. The augmented posterior of the model conditionally
on $\Gamma$ can be written in the form:

\begin{equation}
\begin{array}{c}
p(\alpha,\{\boldsymbol{S}_{t}\},\zeta|\Gamma,X)\propto\\
\prod_{t=1}^{n}g_{\alpha,d}\left(\frac{<X_{t}-\zeta,\boldsymbol{S}_{t}>}{\sigma_{\alpha,\Gamma}(\boldsymbol{S}_{t})},\beta_{\alpha,\Gamma}(\boldsymbol{S}_{t})\right)\sigma_{\alpha,\Gamma}(\boldsymbol{S}_{t})^{-d}\cdot\mathbb{I}(\boldsymbol{S}_{t}\in\mathbb{S}^{d-1})\cdot p(\alpha,\zeta),
\end{array}
\end{equation}

where $p(\alpha,\zeta)$ denotes the prior and the notation $\sigma_{\alpha,\Gamma}(\boldsymbol{S}_{t}),\beta_{\alpha,\Gamma}(\boldsymbol{S}_{t})$
makes explicit the dependence on $\alpha$ and the measure $\Gamma$.
The variables $\boldsymbol{S}_{t}$ are treated as latent variables
with a uniform prior over $\mathbb{S}^{d-1}$ and, provided the measure
$\Gamma$ is approximated with a normal distribution the only unknowns
parameter are $\mu$ and $\omega$ for which we adopt a prior of the
form: $p(\mu,\omega)\propto\omega^{-1}.$ Then, from (13) and (14)
we have:

\begin{equation}
\sigma_{\alpha,\Gamma}(\boldsymbol{S}_{t})^{\alpha}=E_{\boldsymbol{c}\sim\Gamma}|<\boldsymbol{S}_{t},\boldsymbol{c}>|^{\alpha}.
\end{equation}

We notice that:

\begin{equation}
\beta_{\alpha,\Gamma}(\boldsymbol{S}_{t})=\begin{cases}
\begin{array}{c}
\frac{-\iota I_{X}(\boldsymbol{S}_{t})}{\sigma(\boldsymbol{S}_{t})^{\alpha}\tan\frac{\pi\alpha}{2}},\alpha\neq1,\\
\frac{\iota\left\{ I_{X}(2\boldsymbol{S}_{t})-2I_{X}(\boldsymbol{S}_{t})\right\} }{4\sigma(\boldsymbol{S}_{t})\ln\frac{2}{\pi}},\alpha=1.
\end{array}\end{cases}
\end{equation}

\section{MCMC scheme}

The great advantage of (16) is that the measure enters implicitly
through the functions $\sigma_{\alpha,\Gamma}(\boldsymbol{S}_{t}),\beta_{\alpha,\Gamma}(\boldsymbol{S}_{t})$
only. With the Discrete Approximation of the spectral measure, the
computation of functions in (17) and (18) is trivial. When the measure
$\Gamma$ is approximated with a normal distribution the only unknown
parameters are $\mu$ and $\omega$ for which we adopt a prior of
the form: $p(\mu,\omega)\propto\omega^{-1}.$ The expectations can
be approximated as follows:

\begin{equation}
\sigma_{\alpha,\omega}(\boldsymbol{S}_{t})^{\alpha}=M^{-1}\sum_{m=1}^{M}|<\boldsymbol{S}_{t},\boldsymbol{c}_{m}>|^{\alpha},
\end{equation}

where $\boldsymbol{c}_{m}\sim\mathrm{iid}\mathscr{N}(\mu\cdot1_{d},\omega^{2}I_{d}),i=1,...,M$.
The expectation in (19) can be computed easily once we have (4) or
the solution to (7) and $\beta_{\alpha,\omega}(\boldsymbol{S}_{t})$
can be obtained from (18). 

For the Discrete Approximation denote the parameters by $\theta=(\alpha,\zeta,\gamma)$.
For the Normal Approximation the parameter vector is $\theta=(\alpha,\zeta,\mu,\omega)$
and we also have the latent variables $\left\{ \mathbf{s}_{t},t=1,...,n\right\} $
which are absent from the discrete approximation. In the Discrete
Approximation, our prior has the form:

\[
\gamma|\phi\sim\mathscr{N}_{J}(0,\tfrac{\phi^{2}}{\varpi}I),
\]
 subject to the constraints that $\gamma\geq0$ where $\varpi>0$
is a constant and $\phi$ is defined below. To proceed, we write (7)
in the following form:
\begin{quotation}
\label{Linear Char function system}
\begin{equation}
\mathbb{Y}(\bar{\boldsymbol{\tau}}^{K},\boldsymbol{s}^{J})=\mathbb{X}(\bar{\boldsymbol{\tau}}^{K},\boldsymbol{s}^{J})\gamma+\mathbb{U},
\end{equation}

\end{quotation}
where\footnote{Matrices $\mathbb{Y}$ and $\mathbb{X}$ are redefined so that they
contain their real and imaginary parts.} $\mathbb{U}\sim N_{KJ}(0,\phi^{2})$ is an error term that denotes
the deviation between the empirical and theoretical log-characteristic
functions. The bar denotes that the $\boldsymbol{\tau}^{K}$ are fixed.
Combining with (16) we have the posterior resulting from the Discrete
Approximation:

\label{Posterior Discrete}
\begin{multline}
\begin{array}{c}
p(\alpha,\zeta,\{\boldsymbol{S}_{t}\},\Gamma|X)\propto\prod_{t=1}^{n}g_{\alpha,d}\left(\frac{<X_{t}-\zeta,\boldsymbol{S}_{t}>}{\sigma_{\alpha,\Gamma}(\boldsymbol{S}_{t})},\beta_{\alpha,\Gamma}(\boldsymbol{S}_{t})\right)\sigma_{\alpha,\Gamma}(\boldsymbol{S}_{t})^{-d}\cdot p(\alpha,\zeta)\cdot\\
\phi^{-(n+1)}\exp\left\{ -\tfrac{1}{2\phi^{2}}\left[\left(\mathbb{Y}-\mathbb{X}\gamma\right)'\left(\mathbb{Y}-\mathbb{X}\gamma\right)+\varpi\gamma'\gamma\right]\right\} \cdot\mathbb{I}(\gamma\geq0).
\end{array}
\end{multline}

The posterior resulting from the Normal Approximation is the following:

\label{Posterior Normal Approximation }
\begin{multline}
\begin{array}{c}
p(\alpha,\zeta,\{\boldsymbol{S}_{t}\},\mu,\omega|X)\propto\prod_{t=1}^{n}g_{\alpha,d}\left(\frac{<X_{t}-\zeta,\boldsymbol{S}_{t}>}{\sigma_{\alpha,\omega}(\boldsymbol{S}_{t})},\beta_{\alpha,\omega}(\boldsymbol{S}_{t})\right)\sigma_{\alpha,\omega}(\boldsymbol{S}_{t})^{-d}\cdot p(\alpha,\zeta)\cdot\\
\omega^{-(n+1)}\exp\left\{ -\tfrac{1}{2\omega^{2}}\sum_{t=1}^{n}\boldsymbol{S}{}_{t}^{\top}\boldsymbol{S}_{t}\right\} \cdot\prod_{t=1}^{n}\mathbb{I}(\boldsymbol{S}_{t}\in\mathbb{S}^{d-1}).
\end{array}
\end{multline}

We assume throughout that $p(\alpha,\zeta)\propto\mathbb{I}(0<\alpha\leq2)$.
It appears that there are at least three novelties in the formulation
of these posterior distributions. 
\begin{quote}
1. The formal treatment of (7) in the context of (22) which facilitates
considerably the posterior estimation of spectral weights, $\gamma$. 

2. The introduction of latent variables $\left\{ \boldsymbol{S}_{t}\right\} $
to avoid integration over $\mathbb{S}^{d-1}$ in (9).

3. The normal approximation to the measure $\Gamma$ in connection
with (9). 
\end{quote}
The proper prior on $\gamma$ facilitates the regularization of the
troublesome matrix $\mathbb{X}'\mathbb{X}$ through the parameter
$\varpi$, and positive values of $\gamma$ are guaranteed through
the truncation. It is well known that the matrix is ill-conditioned
in the univariate case when $J$$ $ is moderately large, due partly
to the fact that placing optimally the support points is a difficult
problem. See Koutrouvelis (1980, 1981) and Madan and Seneta (1987)
among others. It seems equally difficult to find ``optimal'' placements
in the multivariate case. In this paper we examine sensitivity of
posterior results to alternative configurations of $\boldsymbol{\tau}^{K}$.
Nolan, Panorska and McCulloch (2001) report that the matrix is well-behaved
in the two-dimensional case even with fine grids. Our experience is
similar and implies that one needs to avoid points where the real
and imaginary parts of $I_{X}(\tau)$ have (numerically) the same
value.

Of course, drawing $\gamma$ from the second term in (24) is straightforward.
It is required to draw from:
\begin{quotation}
\label{Proposal of gamma}
\begin{equation}
\begin{array}{c}
\gamma\sim N_{J}(\hat{\gamma},V),\gamma\geq0,\\
\hat{\gamma}=(\mathbb{X}'\mathbb{X}+\varpi I)^{-1}\mathbb{X}'\mathbb{Y},V=\phi^{2}(\mathbb{X}'\mathbb{X}+\varpi I)^{-1}.
\end{array}
\end{equation}

\end{quotation}
This proposal is accepted with certain probability given by the first
term of (24) to maintain the correct posterior distribution. Our MCMC
algorithms (whose details are presented in Appendix A) are as follows.
\begin{quote}
\pagebreak{}$\mathtt{MCMC-Discrete\:Approximation}$\end{quote}
\begin{quotation}
i. Draw $\gamma=\gamma^{*}$ using the normal linear model in (24).
Denote the normal proposal by $q_{\gamma}(\gamma)$. 

ii. Compute (19) and (20) and therefore (22).

iii. Accept the draw with certain probability. 

iv. Update $\theta$ using (22) with given $\boldsymbol{s}_{t}=\bar{\boldsymbol{s}}_{t}$
using proposal $q_{\theta}(\theta).$ Accept the candidate with certain
probability.

v. Update $\boldsymbol{s}_{t}$.\end{quotation}
\begin{quote}
$\mathtt{MCMC-Normal\:Approximation}$\end{quote}
\begin{quotation}
i. Propose $\boldsymbol{S}_{t}^{*}\sim\mathscr{N}_{d}(\mu\cdot1_{d},\omega^{2}I)$.
Denote the proposal by $q(\boldsymbol{s}_{t}^{*}|\theta,X)$.

ii. Compute (19) and (20).

iii. Accept the draw with certain probability based on (23). 

iv. Draw $\theta^{*}\sim q_{\theta}(\theta)$. Accept the candidate
with certain probability based on (23).
\end{quotation}
There are certain numerical issues to resolve. \emph{First}, the choice
of the number of support points, $J$, is made formally using the
log-marginal likelihood, for the MCMC-DA. \emph{Second}, for the parameters
$\theta$ we use as proposals Student-\emph{t} densities, centered
at the maximum likelihood quantities for $\alpha,\zeta$ derived from
fitting univariate stable symmetric distributions. For parameter $\omega$
we use as proposal: $\frac{\sum_{t=1}^{n}\boldsymbol{S}{}_{t}^{\top}\boldsymbol{S}_{t}}{\omega^{2}}\sim\chi_{n}^{2}.$
The univariate stable symmetric density is computed using McCulloch's
(1998) method. ML estimates of $\zeta$ and a diagonal covariance
matrix times a constant $h$ is used in constructing the proposal
for $\zeta$. For $\alpha$ we take the average of ML estimates with
variance $ $$h$ times the median variance from ML estimates. We
adjust $h$ during the transient or ``burn-in'' phase to get acceptance
rates between 20\% and 30\%.

\emph{Third}, To construct a proposal $q_{\gamma}(\gamma)$ for the
(normalized) spectral measure, we begin with draws from standard uniform,
let the MCMC algorithm run through its transient phase and run it
again for another $S_{o}=50,000$ iterations. Then we use uniform
proposals in intervals of the form $[a,b]$ where $a$ and $b$ are
determined from the 99\% probability intervals during the $S_{o}$
phase. The termination of transient phase is determined using Geweke's
(1992) diagnostics every 10,000 passes by comparing the first and
last 2,500 draws. In artificial samples, depending on the parameters,
we needed 50,000 to 150,000 draws. The results are not reported to
save space but a separate appendix is available on request. All reported
results are based on the final 100,000 draws by thinning every other
tenth draw.

\emph{Fourth}, the number of simulations, $M$, to approximate the
functions $\beta(\mathbf{s})$ and $\sigma(\boldsymbol{s})$ is set
to $M=5,000$ and we check in preliminary numerical work whether this
is sufficient. Depending on the values of $\alpha$ and $\boldsymbol{s}$
reasonable precision can be achieved with $M$ ranging from $500$
to $2,500$. For some computational details see the first paragraph
of Section A1 in the Appendix. 

\emph{Fifth}, in (22) and (6) or (7) we need to pre-select the configuration
$\left\{ \boldsymbol{\bar{\tau}}^{K}\right\} $ where the log-characteristic
function is computed. We can use, again, points in $\mathbb{S}^{d-1}$
as in McCulloch (1994). We opt for

\[
\bar{\tau}^{(i)}\sim\mathscr{N}_{d}(0,I),\:<\tau^{(i)},\tau^{(i)}>=1,i=1,...,K,
\]

instead of a uniform distribution. The reason is that we need to concentrate
such points near the origin (Madan and Seneta, 1987, Koutrouvelis,
1980, 1981, Xu and Knight, 2010, and Yu, 2007). We set $K=10J$ so
we have ten times as many points to evaluate the log-characteristic
functions than the number of spectral weights. \emph{Sixth}$ $, the
prior parameter $\varpi=0.01$ which produces a sufficiently diffuse
prior for the spectral weights although we also examine alternative
values for this parameter.

\section{Empirical application}

We consider ten currencies against the US dollar over the period July
3 1996 to May 21 2012 (see Tsionas, 2012). The currencies are Canadian
dollar, Euro, Japanese yen, British pound, Swiss franc, Australian
dollar, Hong-Kong dollar, New Zealand dollar, South Korean won and
Mexican peso. The data is daily and was converted to log differences.
The data are filtered using an AR(1)-GARCH(1,1) model. MCMC is implemented
using a preliminary, transient phase of length 20,000. Then we take
another 100,000 draws from the posteriors of Discrete and Normal Approximations.
The number of support points $J$ for each $\boldsymbol{s}_{t}$ is
determined by running different MCMC chains and computing the approximate
log-marginal likelihood using the method of Lewis and Raftery (1997)
and DiCiccio et al. (1997). For computational details see paragraph
3 in Section A1 of the Appendix.

Student-\emph{t} proposal densities (with 10 degrees of freedom) were
tuned to provide acceptance rates between 20\% and 30\% for the latent
variables. From the results in Table 1 it turns out that Bayes factors
favor $J=20$ points for several values of $\varpi$ so we choose
this value to proceed further with Bayesian analysis with $\varpi=0.01$.
The matrix $\mathbb{X}^{\top}\mathbb{X}$ was not found singular in
all but exceptional cases where approximately (numerically) grids
were generated. In such cases the parameter $\varpi$ resolves the
problem. This computational experience is consistent with the results
in Nolan, Panorska and McCulloch (2001). In their paper they mention
that one needs to scale the data by the median of $|X_{t}|$. Similar
observations were made by Meerschaert and Scheffler (1999) and Tsionas
(2012b). Here we followed the same approach.

\subsubsection*{\newpage{}Table 1. Bayes factors relative to $J=5$}
\begin{quotation}
\begin{tabular}{|c|c|c|c|c|c|c|}
\hline 
$J\rightarrow$ & 10 & 15 & 20 & 30 & 40 & 50\tabularnewline
\hline 
\hline 
$\varpi=10^{-4}$ & 27.12 & 33.87 & 71.23 & 40.01 & 22.1 & 12.67\tabularnewline
\hline 
$\varpi=10^{-3}$ & 29.03 & 35.55 & 81.59 & 32.33 & 10.01 & 6.50\tabularnewline
\hline 
$\varpi=0.01$ & 32.33 & 37.10 & 82.03 & 20.93 & 9.12 & 4.33\tabularnewline
\hline 
$\varpi=0.1$ & 31.44 & 40.32 & 95.44 & 14.32 & 1.83 & 2.16\tabularnewline
\hline 
\end{tabular}
\end{quotation}
In panel (a) of Figure 1 we report marginal posterior densities of
$\alpha$ using the Discrete and Normal approximation for a fixed
configuration $\boldsymbol{\tau}^{K}$. Sensitivity of marginal posteriors
with respect to ten alternative configurations\footnote{Autocorrelation of parameter draws is non-trivial as expected. The
maximum autocorrelation ranged from 0.20 to 0.50 at lag 50 for the
latent variables but was significantly lower for the structural parameters
$\theta$ ranging between 0.10 and 0.30. So there is enough evidence
that the chains mix well. } is examined in panel (b). Marginal posteriors for different values
of $K$ are given in panel (c) up to $K=10J$. Clearly, as $K$ increases
marginal posteriors of $\alpha$ behave in the same way. 

Finally, in panel (d) we report posterior means of the Discrete Approximation
to the spectral measure (thick line) and typical posterior means from
the Normal Approximation under various configurations of $\boldsymbol{\tau}^{K}$.
In fact, the plot underestimates the Normal's ability to approximate
closely the discrete measure (which we take as the ``true'' measure
or its the best approximation) because the discrete measure also changes
with the configuration. Overall, the results are quite robust reasonable
and it is, indeed, encouraging that the Normal Approximation behaves
so well provided, of course, that the multivariate characteristic
function is evaluated at a large enough number of points. Moreover,
the Bayes factor resolves successfully the problem of selecting the
number of support points, $J$, for the latent $\boldsymbol{s}_{t}$.

\subsubsection*{Figure 1. Empirical results for multivariate stable distribution,
exchange rate data}

\includegraphics[scale=0.75]{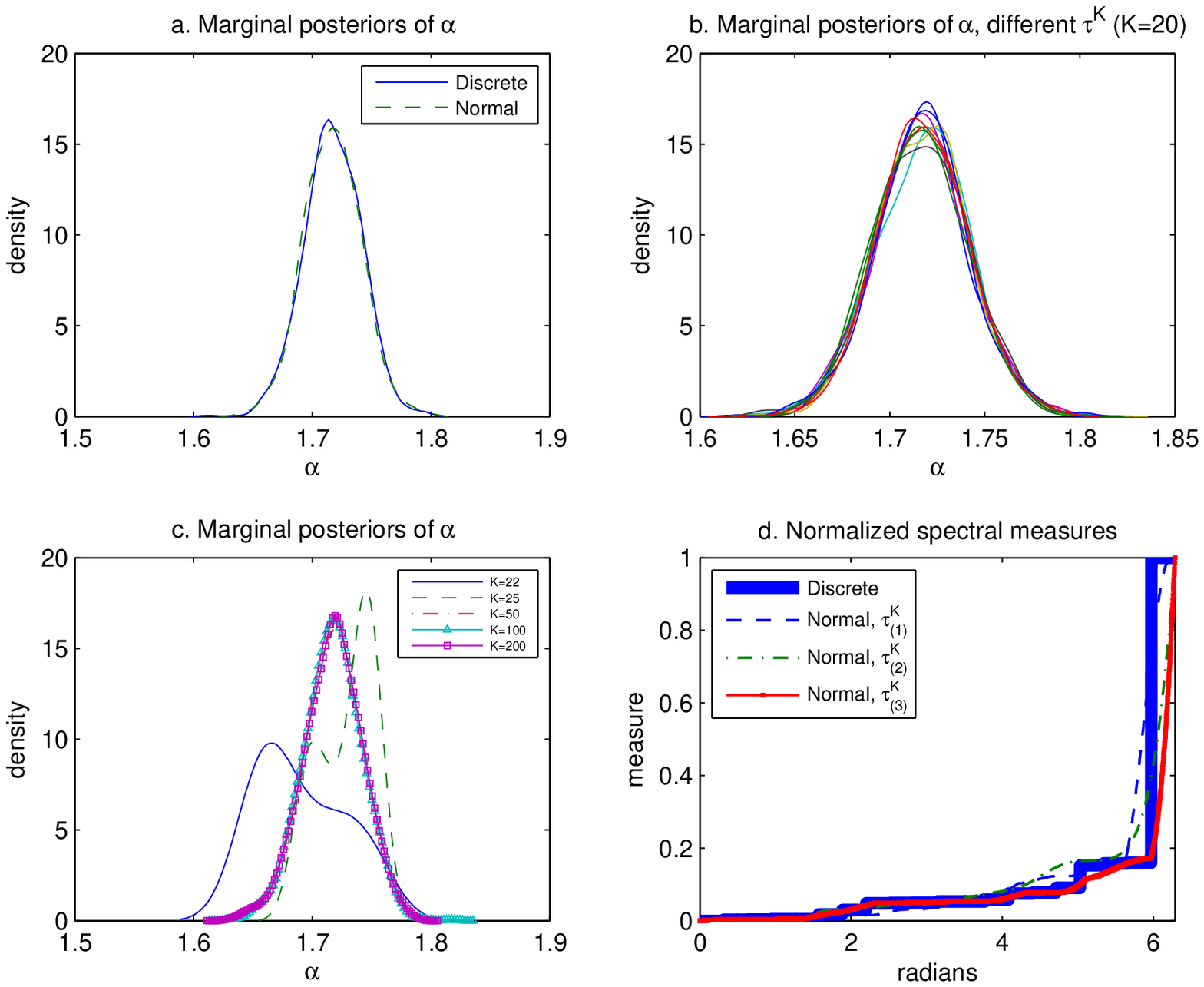}

To study the association between multivariate stable random variables,
the role of the spectral (Levy) measure ($\Gamma$) has been found
critical (Mittnik, Rachev and Rüschendorf, 1999). The dependence-at-the
tails function can be estimated non-parametrically or using the posterior
mean of the Levy measure which has been computed. Technical details
on computing the Mittnik, Rachev and Rüschendorf (1999) function,
$m(x)$, are too many to reproduce here, so we refer instead to their
paper. Specifically, the dependence function is defined in their (2.7)
using (2.6) with the spectral measure involved directly in (2.7) and
(2.4). As they mention: ``If an explicit parametric model is assumed,
it would be more natural and efficient to use the rank-order process
or the dependence function itself'' (p. 184) which is precisely what
we do here.

Our empirical results are reported in Table 2. We report posterior
means of dependence at the tails. The tail is defined as the upper
5\% percentile. Each dependence measure is the posterior mean from
MCMC simulation. The tail dependence measures that include zero in
the Bayes HPD are not reported. Diagonal elements are not reported.

\subsubsection*{\newpage{}Table 2. Empirical results: Tail-Dependence measures}

{\scriptsize{}We report the Mittnik, Rachev and Rüschendorf (1999)
dependence measure at the tails. The tail is defined as the upper
5\% percentile. Each dependence measure is the posterior mean from
MCMC simulation. Tail dependence measures that include zero in the
Bayes HPD are not reported. Diagonal elements are not reported.}{\scriptsize \par}

\begin{tabular}{|c|c|c|c|c|c|c|c|c|c|c|}
\hline 
 & CAD & EUR & JPY & GBP & CHF & AUD & HKD & NZD & KRW & MXN\tabularnewline
\hline 
\hline 
CAD &  & 0.625 &  & 0.742 &  &  &  &  &  & \tabularnewline
\hline 
EUR &  &  & 0.645 & 0.815 & 0.713 & 0.613 & 0.511 & 0.662 & 0.772 & 0.341\tabularnewline
\hline 
JPY &  &  &  & 0.740 & 0.855 & 0.410 &  &  & 0.815 & \tabularnewline
\hline 
GBP &  &  &  &  & 0.603 &  & 0.722 & 0.825 & 0.649 & 0.328\tabularnewline
\hline 
CHF &  &  &  &  &  & 0.515 &  &  & 0.501 & \tabularnewline
\hline 
AUD & 0.610 &  &  &  &  &  &  & 0.717 &  & 0.332\tabularnewline
\hline 
HKD &  &  &  &  &  &  &  &  & 0.423 & \tabularnewline
\hline 
NZD &  &  &  &  &  &  &  &  &  & 0.756\tabularnewline
\hline 
KRW &  &  &  &  &  &  &  &  &  & 0.336\tabularnewline
\hline 
MXN &  &  &  &  &  &  &  &  &  & \tabularnewline
\hline 
\end{tabular}

\medskip{}
Dependence at the tails is, obviously, quite large with the Euro associated
with most currencies followed by the GBP and JPY. Non-parametric measures
computed as in Mittnik, Rachev and Rüschendorf (1999) are somewhat
different, showing that relying explicitly on multivariate stability
delivers some gains in terms of efficiency assuming, of course, the
model is a better description of reality. To our knowledge this is
the first application of the tail-dependence measure provided an explicit
Levy measure $\Gamma$ is used. This measure is computed here as the
posterior mean from Bayes MCMC simulation. 

Given the empirical results it does not appear possible to remove
any currency from the multivariate vector due to its weak dependence
or no dependence at all to other currencies. This is despite the fact
that we have allowed for an AR(1)-GARCH(1,1) scheme. Removing GARCH
effects which are prevalent in many financial time series is essential
in order to satisfy, at least approximately, the i.i.d. assumption
involved in the analysis of multivariate stable distributions.

\section*{Concluding remarks}

In this paper we break new ground in the treatment of multivariate
stable distributions along the following lines. \emph{First}, we propose
a normal approximation to their spectral measure that seems to work
very well in practice. \emph{Second}, we propose efficient MCMC techniques
by introducing appropriate latent variables. These are estimated from
the data along with the spectral measure. \emph{Third}, in connection
with the important per se discrete approximation of the measure, we
estimate it in the context of the simple normal linear model based
on the log-characteristic function. The normal approximation reduces
considerably the computational burden without sacrificing, as it seems,
the quality of the approximation to the benchmark provided by the
discretization of the spectral measure. The fact that it works well
in a data set with ten variables and almost 4,000 observations is
quite encouraging in terms of applications of multivariate stable
distributions.

Posterior inferences seem to be quite robust with respect to the configurations
of $\boldsymbol{\tau}^{K}$. Provided these are normally distributed
over the unit sphere in $\mathbb{R}^{d}$,  our MCMC schemes mix well
with respect to the structural parameters and latent $\boldsymbol{s}^{J}$.

\section*{References}

\noindent Abdul-Hamid, H., and J.P. Nolan, 1998, Multivariate stable
densities as functions of one-dimensional projections, Journal of
Multivariate Analysis 67, 80-89.

Bazant, Z.P., and B.H. Oh, 1986, Efficient numerical integration on
the surface of a sphere, Zeitschrift für Angewandte Mathematik und
Mechanik 1, 37-49. 

Belisle, C., H. Romeijn, and R. Smith, R., 1993, Hit and run algorithms
for generating multivariate distributions, Mathematics of Operations
Research 18, 255-266. 

Broda, S.A., Haas, M., Krause, J., Paolella, M.S., and S.C. Steude,
2012, Stable mixture GARCH models, Journal of Econometrics, in press. 

Buckle, D.J., 1995, Bayesian inference for stable distributions, Journal
of the American Statistical Association 90, 605-613. 

Byczkowski, T., J.P. Nolan, and B. Rajput, 1993, Approximation of
multidimensional stable distributions, Journal of Multivariate Analysis
46, 13-31. 

Cambanis, S., and G. Miller, 1981, Linear problems in \emph{p}th order
and stable processes, SIAM Journal of Applied Mathematics 41, 43-69. 

Cheng, B.N., and S.T. Rachev, 1995, Multivariate stable futures prices,
Mathematical Finance 5, 133-153. 

Chib,S. and E. Greenberg, 1995, Understanding the Metropolis-Hastings
algorithm, The American Statistician 49, 327-335.

DiCiccio, T. J., R.E. Kass, A. Raftery, Adrian, L. Wasserman, 1997,
Computing Bayes Factors by Combining Simulation and Asymptotic Approximations
92, 903-915.

Dominicy, Y., and D. Veredas, 2012, The method of simulated quantiles,
Journal of Econometrics, in press. 

Garcia, R., E. Renault, and D. Veredas, 2011, Estimation of stable
distributions by indirect inference, Journal of Econometrics 161,
325-337. 

Geweke, J.. 1992, Evaluating the accuracy of sampling based approaches
to the calculation of posterior moments, in J.O. Berger, et al. (eds.),
Bayesian Statistics, Vol. 4, pp. 169-194. Oxford: Oxford University
Press.

Hallin, M., Swan, Y., Verdebout, T., and D. Veredas, 2012, One-step
R-estimation in linear models with stable errors, Journal of Econometrics,
in press. 

Koutrouvelis, I. A., 1980, Regression-type estimation of the parameters
of stable laws, Journal of the American Statistical Association 75,
918\textendash 928. 

Koutrouvelis, I. A., 1981, An iterative procedure for estimation of
the parameters of stable laws, Communications in Statististcs - Simulation
and Computation 10, 17\textendash 28. 

Lewis, S.M. and A.E. Raftery, 1997, Estimating Bayes' factors via
posterior simulation with the Laplace- Metropolis estimator, Journal
of the American Statistical Association, 92, 648-655.

Lombardi, M. J., and D. Veredas, 2007, Indirect estimation of elliptical
stable distributions, Computational Statistics and Data Analysis 53,
2309-2324. 

Madan, D.B., and E. Seneta, 1987, Simulation of estimates using the
empirical characteristic function, International Statistical Review
55, 153-161. 

Matsui, M., and A. Takemura, 2009, Integral representations of one-dimensional
projections for multivariate stable densities, Journal of Multivariate
Analysis 100, 334-344.

McCulloch, J. H. , 1994, Estimation of bivariate stable spectral densities,
Technical Report, Department of Economics, Ohio State University. 

McCulloch, J. H., 1998, Numerical Approximation of the Symmetric Stable
Distribution and Density, in R. Adler, R. Feldman, and M. Taqqu (eds),
A practical guide to heavy tails: Statistical techniques for analyzing
heavy tailed data, Boston, Birkhauser, 489-500. 

McCulloch, J. H. , 2000, Estimation of the bivariate stable spectral
representation by the projection method, Computational Economics 16,
47-62.

Meerschaert, M.M., and H.-P. Scheffler, 1999, Moment estimator for
random vectors with heavy tails, Journal of Multivariate Analysis
71, 145-159.

Mittnik, S., S.T. Rachev and L. Rüschendorf, 1999, Test of association
between multivariate stable vectors, Mathematical and Computer Modelling
29 (10-12), 181-195.

Nagaev, A., 2000, On non-parametric estimation of the Poisson spectral
measure of a stable law, Journal of Mathematical Sciences 106, 2854-2859. 

Nolan, J., 1997, Numerical calculation of stable densities and distribution
functions, Communications in Statistics \textendash{} Stochastic Models
13, 759-774. 

Nolan, J. P., 1998, Multivariate stable distributions: approximation,
estimation, simulation and identification. In R. J. Adler, R. E. Feldman,
and M. S. Taqqu (Eds.), A Practical Guide to Heavy Tails: Statistical
techniques for analyzing heavy tailed data, pp. 509-526. Boston: Birkhauser. 

Nolan, J. P., A. Panorska, and J. H. McCulloch, 2001, Estimation of
stable spectral measures, Mathematical and Computer Modelling 34,
1113-1122. 

Nolan, J. P. and B. Rajput, 1995, Calculation of multidimensional
stable densities, Communications in Statistics - Simulation 24, 551-556. 

Ogata, H., 2012, Estimation for multivariate stable distributions
with generalized empirical likelihood, Journal of Econometrics, in
press. 

Paulauskas, V. I., 1976, Some remarks on multivariate stable distributions.
Journal of Multivariate Analysis 6, 356-368. 

Pourahmadi, M., 1987, Some properties of empirical characteristic
functions viewed as harmonizable processes, Journal of Statistical
Planning and Inference 17, 345-359. 

Press, S. J., 1972, Estimation in univariate and multivariate stable
distributions, Journal of the American Statistical Association 67,
842-846. 

Ravishanker, N., and Z. Qiou, 1999, Monte Carlo EM estimation for
multivariate stable distributions, Statistics \& Probability Letters
45, 335-340.

Roose, D., and E. De Doncker, 1981, Automatic integration over a sphere,
Journal of Computational and Applied Mathematics 7, 203-224. 

Samorodnitsky, G., and M. S. Taqqu, 1994, Stable non-Gaussian random
processes, Chapman and Hall, New York. 

Stroud, A., 1971, Approximate Calculation of Multiple Integrals, Prentice
Hall. 

Tierney, L., 1994, Markov chains for exploring posterior distributions,
Annals of Statistics 22 (4), 1701-28.

Tsionas, E.G., 1999, Monte Carlo inference in econometric models with
symmetric stable disturbances, Journal of Econometrics 88, 365-401. 

Tsionas, E.G., 2012, Estimating tail indices and principal directions
easily, Statistics \& Probability Letters 82, 1986-89. 

Zolotarev, V.M., 1986, One-dimensional stable distributions, AMS Translations
of Mathematical Monographs, vol. 65, AMS, Providence, Long Island. 

Xu, D., and J. Knight, 2010, Continuous empirical characteristic function
estimation of mixtures of normal parameters, Econometric Reviews 30,
25-50. 

Yu, J., 2007, Empirical characteristic function and its applications,
Econometric Reviews 23, 93-123. 

\renewcommand{\theequation}{A.\arabic{equation}} 
\setcounter{equation}{0}

\appendix

\section*{Appendix A. Computational details}

\subsection*{A1. Computation of skewness coefficients $\sigma(\boldsymbol{s})$}

\subsection*{1.}

See Bazant and Oh (1986) and Roose and De Doncker (1981). See also
Stroud (1971) for the multidimensional case since the emphasis in
Roose and de Doncker (1981) and Bazant and Oh (1986) is on $\mathbb{S}^{2}$.
Related Fortran code in Stroud (1971) is SPHERE\_05\_ND. In $\mathbb{S}^{2}$
specialized techniques were compared against a hit-and-run algorithm
to generate points uniformly distributed over the spheres. In higher
dimensions we compared using Stroud's (1971) rules. The computational
cost of hit-and run algorithm (Belisle et al, 1993) is trivial even
in high dimensional spaces.  In dimensions up to 20 we have found
that $M$ in the range 500-2,500 provides acceptable results.

\subsection*{2.}

Next, we provide computational experience with the integral $\sigma(\boldsymbol{s})^{\alpha}=\int_{\mathbb{S}^{d-1}}|<\boldsymbol{s},\boldsymbol{t}>|^{\alpha}\Gamma(d\boldsymbol{t})$
when the spectral measure corresponds to $\mathscr{N}_{d}(0,I)$.
The question is, therefore, how to approximate this multivariate integral
accurately and efficiently. The particular measure is of importance
since we use it to approximate the unknown measure. The integral is
approximated as $\sigma(\boldsymbol{s})^{\alpha}\simeq M^{-1}\sum_{m=1}^{M}|<\boldsymbol{s},\boldsymbol{t}_{m}>|^{\alpha}$
where $t_{1},...,t_{M}\sim\mathrm{iid}\mathscr{N}_{d}(0,I)$ . To
assess the approximation we select 1,000 points $\boldsymbol{s}$
\emph{uniformly distributed }over the unit sphere $\mathbb{S}^{d-1}$
and we report the median absolute errors for different values of $d,\alpha$
and $M$. The ``exact'' value of $\sigma(\boldsymbol{s})^{\alpha}$
is obtained using $M=10^{6}$ draws. The ``exact'' value is biased
as the support of $\Gamma$ is $\mathbb{S}^{d-1}$ but it is simulation-consistent.
Draws uniformly distributed in the unit sphere are obtained efficiently
using the hit-and-run algorithm of Belisle et al (1993). The algorithm
is implemented by running 11,000 iterations and keeping the last 1,000.
Convergence was (successfully) tested using Geweke's (1992) diagnostic.
The results are reported in Table A1.

\subsubsection*{\newpage{}Table A1. Median absolute errors of approximation.}
\begin{quotation}
\begin{tabular}{|c|c|c|c|c|c|}
\hline 
 & $M$ & $\alpha=1.1$ & $\alpha=1.5$ & $\alpha=1.75$ & $\alpha=1.9$\tabularnewline
\hline 
\hline 
$d=5$ & 10 & 0.049 & 0.048 & 0.044 & 0.043\tabularnewline
\hline 
 & 100 & 0.018 & 0.017 & 0.016 & 0.015\tabularnewline
\hline 
 & 500 & 0.007 & 0.007 & 0.007 & 0.007\tabularnewline
\hline 
 & 1,000 & 0.005 & 0.005 & 0.005 & 0.005\tabularnewline
\hline 
 & 5,000 & 0.002 & 0.002 & 0.002 & 0.002\tabularnewline
\hline 
$d=10$ & 10 & 0.031 & 0.028 & 0.026 & 0.023\tabularnewline
\hline 
 & 100 & 0.009 & 0.007 & 0.006 & 0.006\tabularnewline
\hline 
 & 500 & 0.006 & 0.005 & 0.005 & 0.005\tabularnewline
\hline 
 & 1,000 & 0.003 & 0.003 & 0.003 & 0.003\tabularnewline
\hline 
 & 5,000 & 0.002 & 0.001 & 0.001 & 0.001\tabularnewline
\hline 
$d=50$ & 10 & 0.020 & 0.012 & 0.008 & 0.007\tabularnewline
\hline 
 & 100 & 0.006 & 0.004 & 0.003 & 0.002\tabularnewline
\hline 
 & 500 & 0.002 & 0.001 & 0.0009 & 0.0008\tabularnewline
\hline 
 & 1,000 & 0.001 & 0.001 & 0.0007 & 0.0006\tabularnewline
\hline 
 & 5,000 & 0.0006 & 0.0004 & 0.0003 & 0.0002\tabularnewline
\hline 
$d=100$ & 10 & 0.009 & 0.004 & 0.003 & 0.002\tabularnewline
\hline 
 & 100 & 0.003 & 0.001 & 0.001 & 0.001\tabularnewline
\hline 
 & 500 & 0.002 & 0.0007 & 0.0005 & 0.0004\tabularnewline
\hline 
 & 1,000 & 0.001 & 0.0006 & 0.0004 & 0.0003\tabularnewline
\hline 
 & 5,000 & 0.0005 & 0.002 & 0.002 & 0.001\tabularnewline
\hline 
\end{tabular}
\end{quotation}
The results are encouraging since for large $d$ even $10$ or $100$
simulations are sufficient to get the integral with sufficient accuracy.
For $d=10$ we need $M=500$ and $5,000$ seems to be quite sufficient.
Since $\sigma(\boldsymbol{s})$ and $\beta(\boldsymbol{s})$ are only
needed in the context of MCMC for acceptance Metropolis-Hastings probabilities
the accuracy reported here is sufficient.

\subsection*{3.}

In (23) the troublesome part in computing the log marginal likelihood
is integration with respect to $\boldsymbol{s}_{t}$. Since the posterior
is evaluated only at the mean, it is relatively easy to integrate
term-by-term using (9) and integration over the sphere (Bazant and
Oh, 1986 and mainly Stroud, 1971). To implement the method of DiCiccio
et al. (1997) we use ``volume correction'' by truncating the MCMC
draws by 5\% in the tails when approximating the posterior at the
mean using a multivariate normal kernel. The dimensionality of the
parameter space increases mainly due to $\zeta\in\mathbb{R}^{d}$.
We did not obtain results significantly different from those reported
in Table 1 when we fixed these parameters to the median of log-differenced
exchange rates, which are close to zero.

\subsection*{A2. Updating support points of the spectral measure}

It is well known that the selection of points is a difficult problem
even in the univariate case. See Madan and Seneta (1987), Koutrouvelis
(1980, 1981), Pourahmadi (1987), Xu and Knight (2010) and Yu (2007).
Generally, a full grid is not optimal.For the Discrete Approximation
the support points are $\boldsymbol{s}^{J}=[s^{(1)},...,s^{(J)}]^{\top}$
where $s^{(j)}\in\mathbb{S}^{d-1}$, for $j=1,...,J$. As a proposal
we use $s^{(j)}\sim\mathscr{N}_{d}(0,hI)|<s^{(j)},s^{(j)}>=1$ where
$h>0$ is a tuning constant. The choice $h=0.5$ performed slightly
better than $h=1$. To improve the mixing of MCMC, given that this
proposal performed rather well we decided to replace the Metropolis-Hastings
step with the so called accept-reject Metropolis-Hastings algorithm
(Tierney, 1994, section 2.3.4, Chib and Greenberg, 1995, section 6.1).
If the algorithm is currently at $\boldsymbol{s}_{o}^{J}$ and a candidate
or proposed move is to $\boldsymbol{s}_{*}^{J}$, define $D=\{\boldsymbol{s}|p(\boldsymbol{s}|\theta,X)\leq cq(\boldsymbol{s})$
as the set where the posterior conditional is dominated by the proposal
$cq(\boldsymbol{s})$. Then the acceptance probability:

\begin{align*}
a(\boldsymbol{s}_{o}^{J},\boldsymbol{s}_{*}^{J}) & =\begin{cases}
\begin{array}{c}
1,\boldsymbol{\:s}_{o}^{J}\in D,\\
\frac{cq(\boldsymbol{s}_{o}^{J})}{p(\boldsymbol{s}_{o}^{J}|\theta,X)},\\
\frac{p(\boldsymbol{s}_{*}^{J}|\theta,X)q(\boldsymbol{s}_{0}^{J})}{p(\boldsymbol{s}_{o}^{J}|\theta,X)q(\boldsymbol{s}_{*}^{J})},\boldsymbol{\:s}_{o}^{J}\notin D,\boldsymbol{s}_{*}^{J}\notin D,
\end{array} & \boldsymbol{s}_{o}^{J}\notin D,\boldsymbol{s}_{*}^{J}\in D\end{cases}
\end{align*}

where $q(\boldsymbol{s}^{J})$ denotes the normal proposal distribution
and $p(\boldsymbol{s}^{J}|\theta,X)$ is the conditional posterior
of $\boldsymbol{s}^{J}$ from the Discrete approximation. To improve
mixing, \emph{the support points are drawn as a group }and by construction
the move from $\boldsymbol{s}_{o}^{J}$ to $\boldsymbol{s}_{*}^{J}$
is certain. In the exchange rate data set the average number of rejections
required to obtain a candidate $\boldsymbol{s}_{*}^{J}$ was 20. The
accept-reject Metropolis-Hastings algorithm performed slightly better
in artificial data with the same dimension $d=10$ and the average
number of rejections was about 15. About 50\% of the time we use an
acceptance test since we have dominance. In artificial data this was
between 60 and 80\% of the time.

For the latent variables $\boldsymbol{S}_{t}$ in the Normal Approximation,
the ``natural'' proposal is a uniform distribution over $\mathbb{S}^{d-1}$.
Other proposals are hard to come by since $g_{\alpha,d}$ involves
computation of a (univariate) integral and the computation of $\sigma_{\alpha,\omega}(\boldsymbol{S}_{t})$.
To facilitate better mixing we adopt the following strategy. Given
the posterior from \ref{Posterior Normal Approximation } we use $M=100$
to obtain an approximate value of $\sigma_{\alpha,\omega}(\boldsymbol{S}_{t})$.
Given the current draw $\boldsymbol{S}_{t}^{o}$ we use one iteration
of the Gauss-Newton method to obtain: 

\[
\boldsymbol{\hat{S}}_{t}=\boldsymbol{S}_{t}^{0}-\left[\nabla^{2}\log p(\boldsymbol{S}_{t}^{o}|\cdot)\right]^{-1}\cdot\nabla\log p(\boldsymbol{S}_{t}^{o}|\cdot),
\]

with numerical derivatives for the gradient and the Hessian. We propose
a move to a candidate 

\[
\boldsymbol{S}_{t}^{*}\sim\mathscr{N}_{d}\left(\boldsymbol{\hat{S}}_{t},-h\cdot\left[\nabla^{2}\log p(\boldsymbol{S}_{t}^{o}|\cdot)\right]^{-1}\right),
\]

which is normalized to lie in $\mathbb{S}^{d-1}$. The candidate is
accepted with probability 

\[
\min\left\{ 1,\frac{p(\boldsymbol{S}_{t}^{*}|\theta,\boldsymbol{s}^{J},X)q(\boldsymbol{S}_{t}^{o})}{p(\boldsymbol{S}_{t}^{o}|\theta,\boldsymbol{s}^{J},X)q(\boldsymbol{S}_{t}^{*})}\right\} ,
\]

where $p(\boldsymbol{S}_{t}|\theta,\boldsymbol{s}^{J},X)$ is the
conditional posterior of $\boldsymbol{S}_{t}$ from \ref{Posterior Normal Approximation }
and $q(\boldsymbol{S}_{t})$ denotes the kernel of the normal proposal
density. The normalizing constant of the proposal is unknown but it
cancels out in the acceptance probability. We use again the accept-reject
Metropolis - Hastings algorithm to guarantee a move from $\boldsymbol{S}_{t}^{o}$
to some new $\boldsymbol{S}_{t}^{*}$. The average number of rejections
was 120 with $h=1$. Using a normal proposal centered directly at
the current $\boldsymbol{S}_{t}^{o}$ was proved impractical since
the average number of rejections was in excess of roughly 5,000. Occasionally,
the proposal centered at $\boldsymbol{\hat{S}}_{t}$ required more
than 10,000 rejections. In such cases we abandon the accept-reject
step, propose a move to $\boldsymbol{S}_{t}^{*}\sim\mathscr{N}_{d}\left(\boldsymbol{S}_{t}^{o},-\left[\nabla^{2}\log p(\boldsymbol{S}_{t}^{o}|\cdot)\right]^{-1}\right)$
and accept with the usual random walk Metropolis - Hastings probability:
$\min\left\{ 1,\frac{p(\boldsymbol{S}_{t}^{*}|\theta,\boldsymbol{s}^{J},X)}{p(\boldsymbol{S}_{t}^{o}|\theta,\boldsymbol{s}^{J},X)}\right\} $.
In the exchange rate data set this occurred in roughly 5\% of draws.
In artificial data the proportion ranged from 0.1\% to 1\% depending
mainly on the value of the characteristic exponent, $\alpha$. 

Regarding updates of $\gamma$ from \ref{Posterior Discrete} the
proposal from \ref{Linear Char function system} was found to work
very well. The acceptance probability is: 

\[
\min\left\{ 1,\frac{p(\gamma^{*}|\alpha,\zeta,\{\boldsymbol{S}_{t}\},X)q(\gamma^{o})}{p(\gamma^{o}|\alpha,\zeta,\{\boldsymbol{S}_{t}\},X)q(\gamma^{*})}\right\} ,
\]

where 

\[
p(\gamma|\alpha,\zeta,\{\boldsymbol{S}_{t}\},X)\propto\prod_{t=1}^{n}g_{\alpha,d}\left(\frac{<X_{t}-\zeta,\boldsymbol{S}_{t}>}{\sigma_{\alpha,\Gamma}(\boldsymbol{S}_{t})},\beta_{\alpha,\Gamma}(\boldsymbol{S}_{t})\right)\sigma_{\alpha,\Gamma}(\boldsymbol{S}_{t})^{-d},
\]

and $q(\gamma)$ denotes again the normal proposal for $\gamma$ as
in \ref{Proposal of gamma}. 

The spectral weights are updated as a group and application of the
accept - reject Metropolis - Hastings algorithm required no more than
1,000 rejections maximum with an average close to 250 and rejection
sampling close to 30\% of the time. The downside is that $n$ univariate
integrals have to be computed to obtain $g_{\alpha,d}(\cdot)$ and
an additional simulation is required to obtain $\sigma(\boldsymbol{S})$.
We have found that this is, indeed, computationally intensive but
the advantage is significant: The MCMC scheme mixes very well contrary
to an additional augmentation by latent variables as in Buckle (1995)
or Tsionas (1999). Our preliminary results with data augmentation
in $d=2$ were somewhat disappointing in the sense that (i) the accept
- reject Metropolis - Hastings takes too long to update successfully,
and (ii) the simple Metropolis - Hastings algorithm takes to stay
far too often. We have of course attempted to tune the algorithms
by a suitable constant in the covariance matrix in \ref{Proposal of gamma}
but it did not seem possible to provide good acceptance rates. We
did not attempt to examine performance in other dimensions although
it is quite possible that other proposals for $\gamma$ might work
better. One such proposal, that we have to leave for future work,
is an importance density drafted along the one - step Gauss - Newton
approach that we described earlier in connection to updating $\boldsymbol{S}_{t}$. 

Proposal distributions for $\alpha$ can be devised in many ways.
Here, we proceed as follows. Suppose $\boldsymbol{\tau}_{(b)}\sim\mathscr{U}(\mathbb{S}^{d-1})$
and consider the projections $y_{t,(b)}=<\boldsymbol{\boldsymbol{\tau}}_{(b)},X_{t}>$,
for $t=1,...,n$ and $b=1,...,B$. For each series $\{y_{t,(b)},t=1,...,n\}$
we estimate the parameter $\hat{\alpha}_{(b)}$ and the location $\hat{\mu}_{(b)}$
of univariate general stable distributions using maximum likelihood
via the FFT. Skewness and scale parameters are functions of $\alpha$
through \ref{sigma(S)} and \ref{beta(S)} but they depend on $\Gamma$
so parameters $\hat{\beta}_{(b)}$ and $\hat{\sigma}_{(b)}$ are estimated
as well. Suppose $\bar{\alpha}=B^{-1}\sum_{b=1}^{B}\hat{\alpha}_{(b)}$
and $\bar{V}_{\alpha}=B^{-1}\sum_{b=1}^{B}(\hat{\alpha}-\bar{\alpha})^{2}$.
Our proposal for $\alpha$ in connection with \ref{Posterior Discrete}
or \ref{Posterior Normal Approximation } is a normal with the indicated
moments. We have used $B=10d=100$ with the exchange rate data. For
the shift parameters $\zeta\in\mathbb{R}^{d}$: (i) During the ``burn
in'' we use a uniform proposal in the interval $(-a,a)^{d}$ and
$a$ is adapted to obtain reasonable acceptance rates (between 20
and 30\%). (ii) During the main phase we keep $a$ fixed.

\section*{A3. An alternative proposal for the spectral measure}

Since the spectral measure, $\Gamma$, is of critical importance in
multivariate stable distributions it is, perhaps, desirable to examine
alternative proposal or importance distributions. From \ref{sigma(S)}
given estimates of $\hat{\alpha}$ and $\hat{\sigma}(\boldsymbol{s})$
and a discrete approximation for $\Gamma$, we have

\[
\hat{\sigma}(\boldsymbol{t})^{\hat{\alpha}}=\sum_{j=1}^{I}\gamma_{j}\cdot|<\boldsymbol{s}_{(j)},\boldsymbol{t}>|^{\alpha},
\]

where $\boldsymbol{s}^{J}=\{\boldsymbol{s}_{(j)},j=1,...,I\}$ denotes
the points of the support corresponding to partition of the unit sphere.
Their number, $I$, is not related to $J$. From this expression we
obtain:

\[
\hat{\sigma}(\boldsymbol{\boldsymbol{t}}_{(i)})^{\hat{\alpha}}=\sum_{j=1}^{I}\gamma_{j}\cdot|<\boldsymbol{t}_{(i)},\boldsymbol{s}_{(j)}>|^{\hat{\alpha}},i=1,...,N,j=1,...,I,
\]

where $\boldsymbol{t}^{N}=\{\boldsymbol{t}_{(1)},...,\boldsymbol{t}_{(N)}\}$
denotes the set set of projections where the scale parameter is estimated.
The idea here is to fix a set of $\boldsymbol{s}_{(j)}$. Then we
know that $<\boldsymbol{t}_{(i)},X_{t}>$ follows a univariate stable
distribution with characteristic exponent $\alpha$ and skewness and
scale given as in \ref{sigma(S)} and \ref{beta(S)}. These can be
estimated using maximum likelihood and the FFT to obtain the log -
likelihood function. The system has the form:

\[
\begin{array}{c}
\boldsymbol{A}\gamma=\boldsymbol{b},\\
\boldsymbol{A}_{(N\times I)}=[a_{ij}],a_{ij}=|<\boldsymbol{t}_{(i)},\boldsymbol{s}_{(j)}>|^{\hat{\alpha}},\\
\boldsymbol{b}_{(N\times I)}=[b_{i}],b_{i}=\hat{\sigma}(\boldsymbol{\boldsymbol{t}}_{(i)})^{\hat{\alpha}},i=1,...,N,j=1,...,I,
\end{array}
\]

and it is somewhat easier to solve than the corresponding system in
\ref{CF system no error} which uses the characteristic function.
More importantly, this system will be solved only once to construct
a proposal for updating $\gamma$ in the context of MCMC. Specifically
we choose $I=100$ points $\boldsymbol{s}_{(j)}\sim\mathcal{\mathscr{\mathrm{iid}N}}_{d}(0,I)\parallel\mathbb{S}^{d-1}$
and $N=10\cdot I=1,000$ points $\mathbf{s}_{(i)}\sim\mathscr{U}(\mathbb{S}^{d-1})$
and we use Bayesian inference with non-negativity constraints on the
$\gamma$, in the system $\boldsymbol{b}=\boldsymbol{A}\gamma+\boldsymbol{u}$
where $\boldsymbol{u}\sim\mathrm{iid}\mathscr{N}_{N}(0,\sigma_{b}^{2}I_{N})$.
We use an adaptive Metropolis - Hastings algorithm (with $\gamma=\delta\odot\delta$
and $\delta\in\mathbb{R}^{I}$ ) to perform the computation with 15,000
passes the first 5,000 of which are discarded. From the remaining
draws we retain only every other tenth and we estimate the mean, $\bar{\gamma}$
and the covariance matrix $\bar{V}$ from the MCMC draws. 

Since the log marginal likelihood of this model can be computed easily
using the Laplace approximation, one can set, \emph{if desired}, $J=I^{*}$
and $I^{*}$ attains the maximum value of the log marginal likelihood.
Here, $J$ is the number of support points for the spectral measure
that we use in the main text. It was made clear that we did not opt
for this choice as it does not fully utilize the information in the
data. However, it is useful in that \emph{it can provide a proposal
distribution for the support points $\boldsymbol{s}_{(j)}$ of the
spectral measure}. For the support points, in particular, this can
be achieved as follows. The MCMC procedure in the linear system is
repeated 100 times with 100 different sets of $\boldsymbol{s}^{I}=\left\{ \boldsymbol{s}_{(j)},j=1,...,I\right\} $
for a fixed set of $\boldsymbol{t}^{N}$. The set that corresponds
to the best value of the log marginal likelihood is $\{\boldsymbol{s}_{*}^{I},\gamma_{*}\}$.
This is repeated for 10,000 alternative sets of $\boldsymbol{t}^{N}$
resulting in a set $\mathcal{\mathscr{G}}_{N}=\{\boldsymbol{s}_{*,(i)}^{I},\gamma_{*,(i)},i=1,...,N\}$.
From this set whose variability is due to the different configurations
of $\boldsymbol{t}^{N}$ we can determine a joint proposal distribution
for $\tilde{\Gamma}=\{\boldsymbol{s}_{(j)},\gamma_{j},j=1,...,I\}$
which characterizes completely the spectral measure, $\Gamma$. The
joint proposal distribution is a uniform for each element of $\tilde{\Gamma}$
whose bounds are determined from the 99\% probability intervals of
$\mathcal{\mathscr{G}}_{N}$ point-wise. 

This well-crafted proposal can be used to update jointly the spectral
weights $\gamma$ and the spectral support points $\{\boldsymbol{s}_{(j)},\gamma_{j},j=1,...,I\}$.
The number of support points, $I$, was set intentionally to a large
number $I=100$ although preliminary investigations revealed that
the optimal $I^{*}$ defined earlier was much lower. Our MCMC scheme
for inference in multivariate stable distributions is performed for
fixed $J$. For each specific value of $J$ for which we implement
MCMC in the main text we have tried two alternatives.
\begin{quotation}
i. We set $J=I$ and repeat the previous analysis so that we have
an ``exact'' proposal for $\tilde{\Gamma}$ that matches exactly
with the spectral weights and support points that we need to update.

ii. We use $I=100$ throughout. For each specific value of $J<I$,
we re-normalize $\mathscr{G}_{N}$ and $\tilde{\Gamma}$ discarding
weights (and the corresponding support points) less than $\epsilon_{I}$
which is determined by the requirement that we have $I$ support points.
Given the support points the weights that are less than $\epsilon_{I}$
are allocated to the retained support points that are closest (in
the $L_{1}-$ norm) to the excluded support points.
\end{quotation}
With the exchange rate data set the two alternatives did not behave
in a qualitatively different way. The posterior estimates of structural
parameters and more importantly of the spectral measure are quite
close. This observation implies that MCMC explores the posterior quite
satisfactorily once we have a ``reasonable'' procedure.

Moreover, we have used a fixed set of projections $\{\tau_{(i)}\}$
which does not correspond to the $\boldsymbol{t}^{N}$ used here,
since we wanted to construct a procedure that can be applied more
generally. We use, again, Tierney's (1994) accept - reject Metropolis
- Hastings. This, again, provides sufficient evidence that our MCMC
explores the posterior quite well. 

In Table A1 we report some statistics relating to the two proposals
we have constructed. For the main parameters of the model we report
the number of rejections in the accept-reject step to obtain a candidate
when dominance holds and (in parentheses) the percentage of draws
for which the accept-reject step was implemented (due to dominance)
from the ``blanketing'' proposal.

\subsubsection*{\newpage{}Table A2. Statistics relating to the performance of proposals}
\begin{quotation}
\begin{tabular}{|c|c|c|c|}
\hline 
 & Proposal, section A2 & Proposal, section A3 & acf, lag 50$^{\star}$\tabularnewline
\hline 
\hline 
$\left(\gamma,\left\{ \boldsymbol{s}_{(j)}\right\} \right)$ & 67 (43.5\%)$^{\dagger}$ & 12 (62.1\%) & 0.21 0.15\tabularnewline
\hline 
$\gamma$ & 25 (50.8\%) & 32 (44.4\%)$^{\ddagger}$ & 0.32 0.37\tabularnewline
\hline 
$\left\{ \boldsymbol{s}_{(j)}\right\} $ & 32 (43.2\%) & 91 (12.3\%)$^{\ddagger}$ & 0.18 0.22\tabularnewline
\hline 
$\left(\alpha,\zeta\right)$ & 15 (25.5\%) & 7 (62.3\%) & 0.25 0.35\tabularnewline
\hline 
$\left\{ \boldsymbol{S}_{t}\right\} $$^{\dagger\dagger}$ & 210 (21.3\%) & 12 (33.7\%) & 0.11 0.07\tabularnewline
\hline 
\end{tabular}

Notes: $^{\dagger}$ In this case we update jointly $\left(\gamma,\left\{ \boldsymbol{s}_{(j)}\right\} \right)$.

$^{\ddagger}$ In this case we update separately $\gamma$ and $\left\{ \boldsymbol{s}_{(j)}\right\} $. 

$^{\dagger\dagger}$ The reported numbers are medians for $t=1,...,n$. 

$^{\star}$ This is autocorrelation at lag 50. The first number refers
to the proposal of section A2 and the second number refers to the
proposal of section A3.
\end{quotation}
Clearly the proposal of this section performs slightly better and
``blankets'' the posterior conditional distributions of the key
parameters better. However, \emph{the important point is that the
proposal of section A2 is quite competitive relative to this well-crafted
proposal.} It seems, therefore, that in practical applications the
overhead of constructing proposal distributions by preliminary fitting
of the projection - dependent scale parameter $\sigma(\boldsymbol{s})$
will, most likely, not be worth the effort. The results reported here
refer to a single data set (the exchange rates) so we cannot generalize
to all empirical applications although this would constitute an important
matter for future research. 
\end{document}